# The properties of isolated Main Galaxies in Sloan Digital Sky Survey Data Release 3 (SDSS DR3)

Deng Xinfa[A,B]  Ma Xinsheng[A]  Luo Chenghong[A]  Zhang Qun[A]  He Ji-zhou[A]

[A]Mathematics and Physics College, Nanchang University, Jiangxi, China, 330047
[B]xinfadeng@163.com

**Abstract**   We have constructed a Main galaxy subsample with redshifts in the range $0.08 \leq z \leq 0.12$ from SDSS Data Release 3 (SDSS DR3), and refer to this subsample as "MGSU"(It contains 67777 galaxies). By cluster analysis, two isolated Main galaxy samples are extracted from this subsample. It turns out that two isolated Main galaxy samples identified at different radii have the same properties. Additionally, we also find that early-type galaxies in isolated Main galaxy samples are fewer than that in close double galaxy sample..

**Keywords**   cosmology－galaxy－cluster analysis

## 1. Introduction

Isolated galaxies, which may have experienced no major interactions in billions of years, are a group of special and rare galaxies in the universe. They are important because they probe the lowest density regions of galaxies. They can serve as an interesting sample for detecting and studying the properties of galaxies, for example, studies of the effects of environment on galaxy morphologies and star formation rates (Adams, Jensen & Stocke 1980; Haynes & Giovanelli 1980; Haynes, Giovanelli & Chincarini 1984; Koopmann & Kenney 1998). In recent years, many properties of isolated galaxies were explored fully (Aars 2003; Pisano & Wilcots 2003; Sauty et al. 2003; Stocke et al. 2004; Varela et al. 2004; Marcum et al. 2004).

According to Karachentseva (1973)'s selection algorithm, a galaxy i with angular diameter $a_i$ is considered isolated if the projected sky separation $x_{i,j}$ between this galaxy and any neighboring galaxy j of angular diameter $a_j$ satisfies the following two criteria: $x_{i,j} \geq 20 \times a_j$; $\frac{1}{4} a_j \leq a_i \leq 4 \times a_j$. Using Zwicky et al. (1968)'s Catalogue of Galaxies & Clusters of Galaxies, Karachentseva compiled her isolated galaxy catalog from work with prints of the Palomar Sky Survey. The original catalog contained 1052 candidate isolated galaxies, which was later reduced to 893 galaxies (Karachentseva 1980).

Allam et al. (2005) implemented a variation on the original Karachentseva criteria, and identified isolated galaxies in Sloan Digital Sky Survey Data Release 1 (SDSS DR1). According to their modified criteria, a galaxy i with a g-band magnitude $g_i$ and g-band

Supported by the National Science Foundation of China (10465003)

Petrosian radius $R_i$ is considered to be isolated if the projected sky separation between this galaxy and any neighboring galaxy j satisfies: $x_{i,j} \geq 40 \times R_j$; $|g_i - g_j| > 3.0$. After all rejections and verifications, 2980 isolated galaxies were extracted from imaged coverage of about 2099 deg$^2$.

Above methods of identifying isolated galaxies mainly are based on the two-dimensional projected sky separation and galaxy diameter. The selected galaxies may not be the most isolated in the three-dimensional space. When we use a catalog of galaxies with redshift (three-dimensional galaxy sample), identifying isolated galaxies only by the projected separation and diameter criteria is not quite appropriate. In this paper, we intend to use three-dimensional cluster analysis (Einasto et al. 1984) and extract isolated Main galaxies from SDSS Data Release 3 (Abazajian et al. 2005). By cluster analysis, the sample can be separated into isolated galaxies, close double and multiple galaxies, galaxy groups or clusters. At larger radii, most galaxies of the sample form different groups or clusters, few galaxies are isolated. These isolated galaxies should be a good sample for studies of three-dimensional isolated galaxies.

Our paper is organized as follows. In section 2, we describe the data to be used. The cluster analysis and selection criteria are discussed in section 3. In section 4, we present basic properties of isolated Main galaxies. Our main results and conclusions are summarized in section 5.

## 2. Data

The Sloan Digital Sky Survey (SDSS) is one of the largest astronomical surveys to date. The completed survey will cover approximately 10000 square degrees. York et al. (2000) provided the technical summary of the SDSS. The SDSS observes galaxies in five photometric bands (u，g，r，i，z) centered at (3540, 4770, 6230, 7630, 9130°A). The imaging camera was described by Gunn et al. (1998), while the photometric system and the photometric calibration of the SDSS imaging data were roughly described by Fukugita et al. (1996), Hogg et al. (2001) and Smith et al. (2002) respectively. Pier et al. (2003) described the methods and algorithms involved in the astrometric calibration of the survey, and present a detailed analysis of the accuracy achieved. Many of the survey properties were discussed in detail in the Early Data Release paper (Stoughton et al. 2002). Galaxy spectroscopic target selection can be implemented by two algorithms. The primary sample (Strauss et al. 2002), referred to here as the MAIN sample, targets galaxies brighter than r < 17.77(r-band apparent Petrosian magnitude). The surface density of such galaxies is about 90 per square degree. This sample has a median redshift of 0.10 and few galaxies beyond z=0.25. The Luminous Red Galaxy (LRG) algorithm (Eisenstein et al. 2001) selects ～12 additional galaxies per square degree, using color-magnitude cuts in g, r, and i to select galaxies to r < 19.5 that are likely to be luminous early-types at redshifts up to ～0.5.

The SDSS has adopted a modified form of the Petrosian (1976) system for galaxy photometry, which is designed to measure a constant fraction of the total light independent of the surface-brightness limit. The Petrosian radius $r_p$ is defined to be the radius where the local surface-brightness averaged in an annulus equals 20 percent of the mean surface-brightness interior to this annulus, i.e.

$$\frac{\int_{0.8r_p}^{1.25r_p} dr 2\pi\, rI(r)/[\pi(1.25^2 - 0.8^2)r^2]}{\int_{0}^{r_p} dr 2\pi\, rI(r)/[\pi r^2]} = 0.2$$

where I(r) is the azimuthally averaged surface-brightness profile. The Petrosian flux $F_p$ is then defined as the total flux within a radius of $2r_p$, $F_p = \int_{0}^{2r_p} 2\pi\, rdrI(r)$. With this definition, the Petrosian flux (magnitude) is about 98 percent of the total flux for an exponential profile and about 80 percent for a de Vaucouleurs profile. The other two Petrosian radii listed in the Photo output, $R_{50}$ and $R_{90}$, are the radii enclosing 50 percent and 90 percent of the Petrosian flux, respectively.

The SDSS sky coverage can be separated into three regions. Two of them are located in the north of the Galactic plane, one region at the celestial equator and another at high declination. The third lies in the south of the Galactic plane, a set of three stripes near the equator. Each of these regions covers a wide range of survey longitude.

In our work, we consider the Main galaxy sample. The data is download from the Catalog Archive Server of SDSS Data Release 3 (Abazajian et al. 2005) by the SDSS SQL Search (with SDSS flag: bestPrimtarget=64) with high-confidence redshifts (Zwarning ≠ 16 and Zstatus ≠ 0, 1 and redshift confidence level: zconf>0.95) (http://www.sdss.org/dr3/coverage/credits.html). From this sample, we select 214795 Main galaxies in redshift region: 0.02≤z≤0.2.

In calculating the distance we use a cosmological model with a matter density $\Omega_0 = 0.3$, cosmological constant $\Omega_\Lambda = 0.7$, Hubble's constant $H_0 = 100h\,km\cdot s^{-1}\cdot Mpc^{-1}$ with h=0.7.

Fig.1 illustrates the distribution of redshift z of galaxies for our Main galaxy sample. The peak of the distribution is at about z=0.08. Because Main galaxy sample is an apparent-magnitude limited sample, the space density of galaxies decreases apparently with growing redshift z. Fig.2 shows the changing of the space density of galaxies with redshift z. This kind of the incompleteness of sample will produce certain influence upon the result of identifying isolated galaxies. The number of isolated galaxies identified in the low density region will be more than that of isolated galaxies identified in the high density region. In order to decrease the effect of radial selection function, we have constructed a subsample in the redshift region $0.08 \leq z \leq 0.12$, and refer to this subsample as "MGSU"(It contains 67777 galaxies).

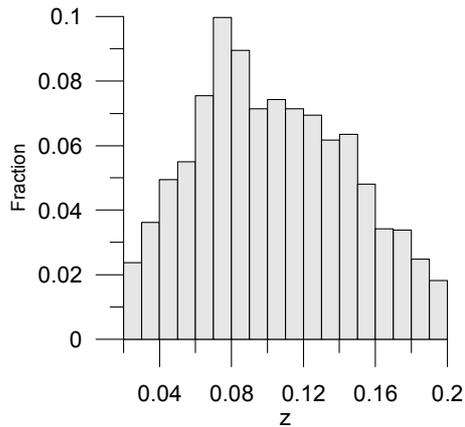
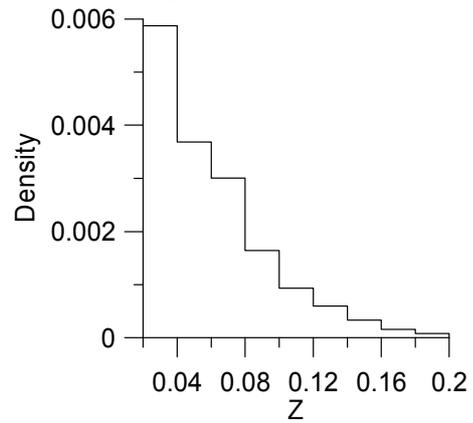

Fig.1 The distribution of redshift z of galaxies for our Main galaxy sample

Fig.2 The distribution of space density of galaxies with redshift z for our Main galaxy sample

## 3. Cluster analysis

Cluster analysis (Einasto et al. 1984), as a general method, has been widely applied to study the geometry of point samples in many fields. The key of this method is to separate the sample into individual systems by an objective, automatic procedure. Let us draw a sphere of radius R around each sample point (in our case, galaxy). If within this sphere there are other galaxies they are considered belonging to the same system. Call these close galaxies "friends". Now draw spheres around new neighbours and continue the procedure using the rule "any friend of my friend is my friend". When no more new neighbours or "friends" can be added, then the procedure stops and a system is identified. As a result, each system consists of either a single, isolated galaxy or a number of galaxies which at least have one neighbour within a distance not exceeding R. Apparently, the number of isolated galaxies decreases with growing dimensionless radii r. We set up the threshold that the number of isolated galaxies is below 10% of total galaxy number in the sample as the condition of producing isolated galaxy sample.

The mean density of galaxies is $\overline{\rho} = N/V$ (N is the number of galaxies contained in the volume V). The radius of the sphere with unit population is $R_0 = (3/4\pi\overline{\rho})^{1/3}$ (For "MGSU", $R_0$ is 5.7863Mpc). In our analysis, we express all distances in dimensionless radii $r = R/R_0$ .

## 4. The basic properties of isolated Main galaxies

When we identify isolated galaxies by cluster analysis, it is important to realize that we do not have any a priori defined neighbourhood radius to identify isolated galaxies. This forces us to consider and analyse a certain range of neighbourhood radii, accordingly the properties of systems forming in sample. At small radii only close double and multiple galaxies, cores of groups and conventional clusters of galaxies will form systems, the most being isolated single galaxies. With a growing neighbourhood radius less dense peripherical regions of groups and clusters will be included into systems. At still larger radii neighbouring systems merge into large, usually elongated units, few galaxies are isolated. Fig.3 and Fig.4 respectively show how the number $N_{sin}$ of isolated single galaxies and the galaxy number $N_{max}$ of the richest system change with growing dimensionless radius r for "MGSU". Fig.4 demonstrates that at radius r=0.9, the galaxy number of the richest system is 2687, and starts a very steep increase with radii r (at r=0.8, it is 244). At this radius, most galaxies of the sample begin to merge into less dense regions of groups and clusters, a large and less dense system is formed. At radius r=1.2, the galaxy number of the richest system reachs 7355, the number of isolated galaxies is 4099. Fig.3 shows that at this radius the change of number $N_{sin}$ of isolated single galaxies gradually becomes even as the radius r increases. Isolated galaxies identified at radii $r \geq 1.2$ can be defined as real isolated galaxies in three-dimensional space. We select the isolated galaxies identified at radii r=1.2 and

r=1.4 as two isolated galaxy samples, respectively refer to Sample1.2 and Sample1.4. Sample1.2 includes 4099 isolated galaxies (6.0% of the total galaxy number in "MGSU"), Sample1.4 includes 2596 isolated galaxies (3.8% of the total galaxy number in "MGSU").

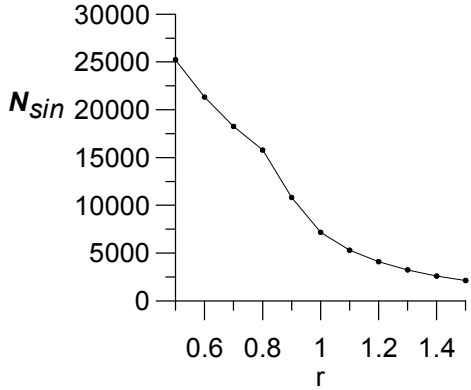 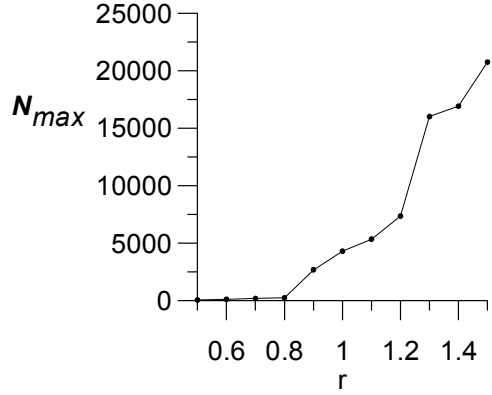

Fig.3　Isolated galaxy number $N_{sin}$ changes with dimensionless radius r.

Fig.4　The galaxy number $N_{max}$ of the richest system changes with dimensionless radius r.

Luminosity, size, and morphological type are the most basic properties of a galaxy. Observed galaxies cover large ranges in these properties, effective radii between 0.1kpc and 50kpc, morphologies changing from pure disk systems to pure ellipsoidal systems. Clearly, the study of the distribution of galaxies with respect to these properties are crucial to our understanding of the formation and evolution of the galaxy population.

Fig.5 shows the histogram of distribution of the luminosities $M_r$ of galaxies for "MGSU", Sample1.2 and Sample1.4. The $M_r$ is the r-band absolute magnitude. For galaxies at z < 0.12, the K-corrections are trivial. So we ignore the K-correction (Blanton et al. 2002). The absolute magnitude $M_r$ is calculated from the r-band apparent magnitude $m_r$ as follows:
$M_r = m_r - 5\log D_L - 25$

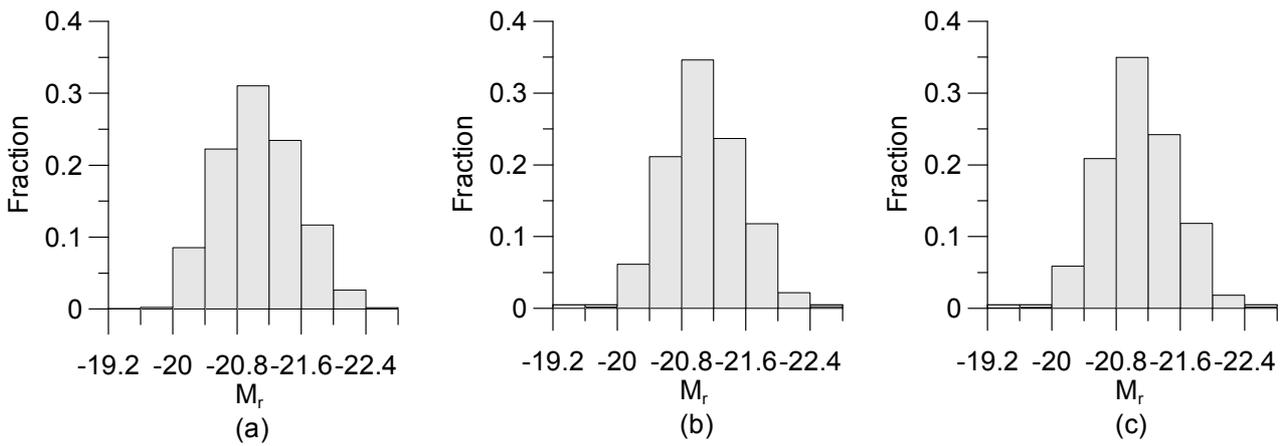

Fig.5　The histogram of distribution of the luminosities $M_r$ of galaxies for "MGSU", Sample1.2 and Sample1.4. (a) for "MGSU", (b) for Sample1.2, (c) for Sample1.4.

$D_L$ is the luminosity distance. We find that the distribution of the luminosities $M_r$ of galaxies for "MGSU", Sample1.2 and Sample1.4 are almost the same.

We select the r-band $R_{50}$ ($R_{50,r}$) as the parameter of galaxy size, and have calculated the concentration index ci=$R_{90}/R_{50}$ which can be used to separate early-type (E, So) galaxies from late-type (Sa, Sb, Sc, Irr) galaxies (Shimasaku et al. 2001). Using about 1500 galaxies with eye-ball classification, Nakamura et al. (2003) confirmed that ci=2.86 separates galaxies at So/a with a completeness of about 0.82 for both late and early types. Fig6 and Fig.7 respectively show the size distribution of galaxies and the concentration index ci distribution of galaxies for "MGSU", Sample1.2 and Sample1.4. In Sample1.2, 22.7%galaxies have the concentration index ci>2.86, in Sample1.4 this proportion is 22.1%, is apparently lower than the proportion (28.3%) in "MGSU". This demonstrates that there are fewer early-type galaxies in isolated galaxy samples.

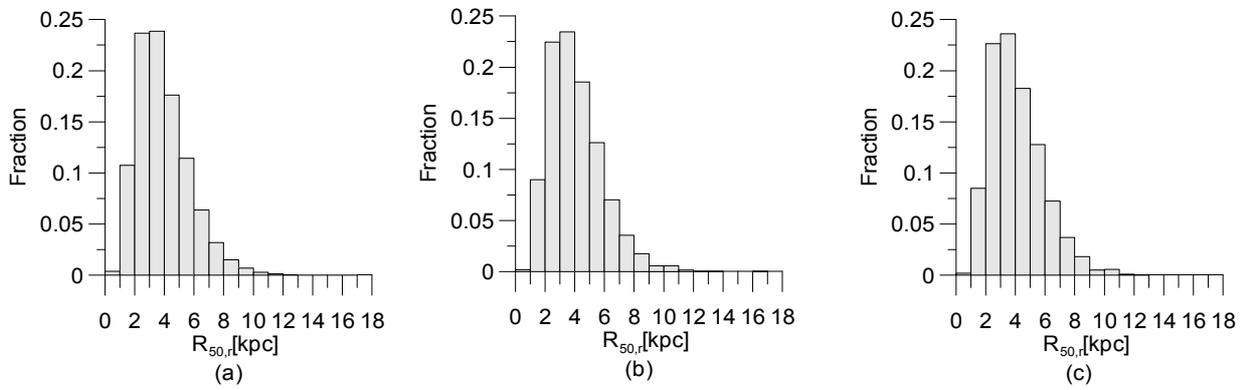

Fig.6  The histogram of the size distribution of galaxies for "MGSU", Sample1.2 and Sample1.4. (a) for "MGSU", (b) for Sample1.2, (c) for Sample1.4.

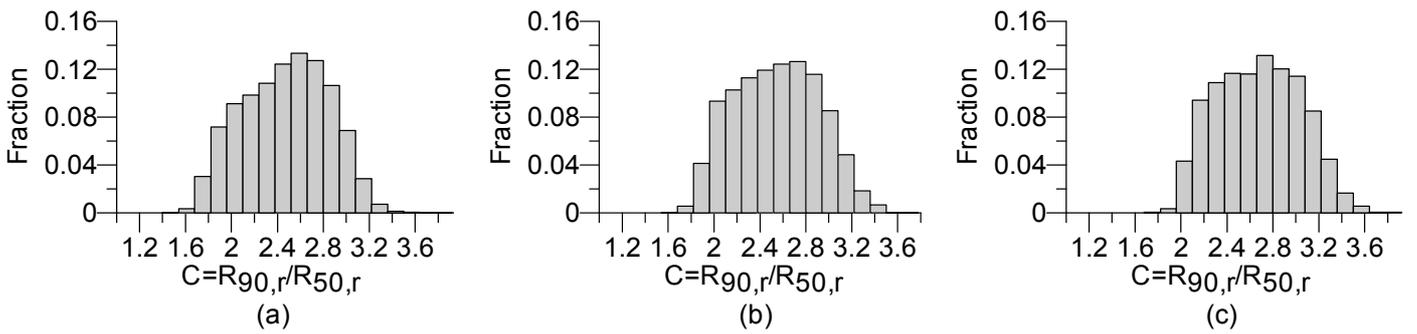

Fig.7  The histogram of the concentration index ci distribution of galaxies for "MGSU", Sample1.2 and Sample1.4. (a) for "MGSU", (b) for Sample1.2, (c) for Sample1.4.

According to above analysis, we find that isolated galaxy samples identified at different radii have the same properties. This demonstrates that isolated galaxies identified at this radius region by cluster analysis have definite basic properties and can be considered good isolated galaxy samples.

At very small radii only close double and multiple galaxies will form systems. Moreover most systems are close double galaxies. In order to compile a real three-dimensional galaxy pair catalog, Deng et al. (2005) extracted close double Main galaxies from SDSS Data Release 3 by cluster analysis. They selected the range of projected separation criteria of most pair samples as the analysis range of neighbourhood radii, and analysed the clustering properties of Main galaxy sample in this neighbourhood radius range: $R = 60\text{kpc} \rightarrow 200\text{kpc}$, in order to find the appropriate neighbourhood radius to identify galaxy pairs. Meanwhile, the distribution of the r-band $R_{90}$ ($R_{90, r}$) of all galaxies for Main galaxy sample was analysed. Finally, they concluded that the neighbourhood radius $R \approx 100\text{kpc}$ can be defined as appropriate neighbourhood radius to identify close galaxy pairs. At neighbourhood radius R=100kpc, we identify 570 close double galaxies in "MGSU" (0.84% of the total galaxy number in "MGSU") and select these close double galaxies as our galaxy pair sample. In Fig.8, Fig.9 and Fig.10, we compare the luminosity $M_r$, size and concentration index $ci$ distributions of galaxies of pair sample identified at radius R=100 kpc with that of isolated galaxy sample identified at dimensionless radius r=1.2 (Sample1.2). It turns out that there are more early-type ($ci>2.86$)galaxies in galaxy pair sample (in isolated galaxy sample, galaxies of concentration index $ci>2.86$ are about 22.7%, while in pair sample this proportion is 37.4%). Additionally, We also notice that on the average there is higher proportion of large galaxies(larger than 4 kpc) in isolated galaxy sample.

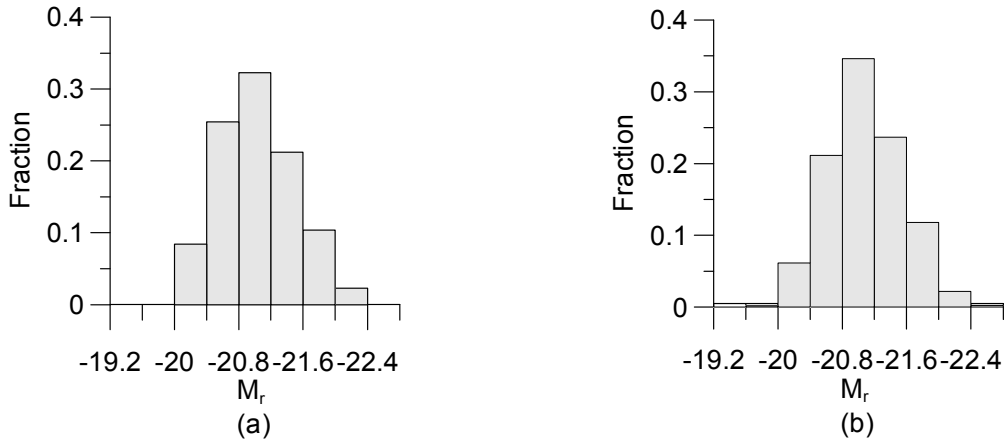

Fig.8    The histogram of the luminosity distribution of galaxies for galaxy pairs identified at neighbourhood radius R=100 kpc and isolated galaxies identified at dimensionless radius r=1.2 (Sample1.2). (a) for galaxy pairs identified at neighbourhood radius R=100 kpc, (b) for Sample1.2.

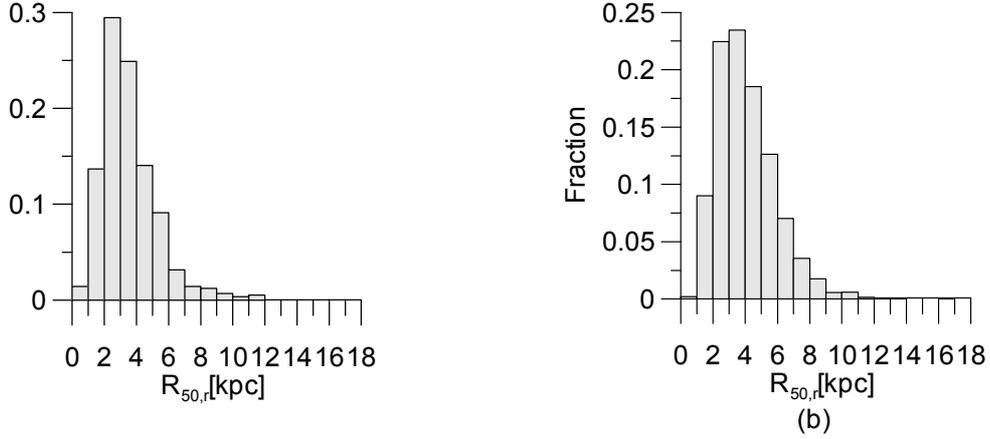

Fig.9 The histogram of the size distribution of galaxies for galaxy pairs identified at neighbourhood radius R=100 kpc and isolated galaxies identified at dimensionless radius r=1.2 (Sample1.2). (a) for galaxy pairs identified at neighbourhood radius R=100 kpc, (b) for Sample1.2.

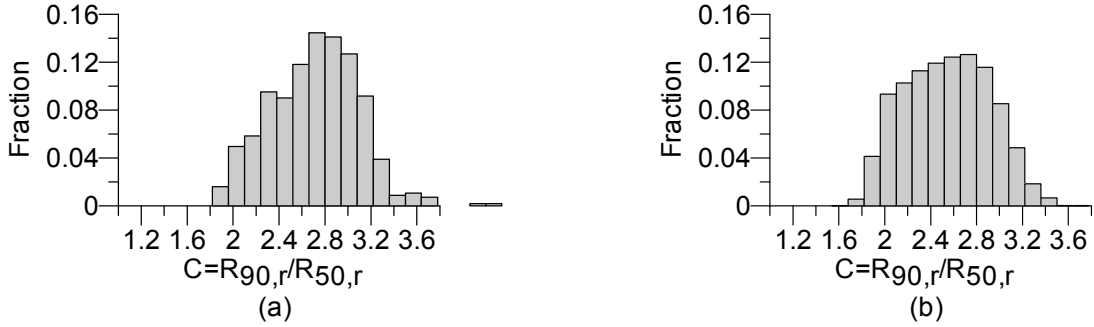

Fig.10 The histogram of the concentration index ci distribution of galaxies for galaxy pairs identified at neighbourhood radius R=100 kpc and isolated galaxies identified at dimensionless radius r=1.2 (Sample1.2). (a) for galaxy pairs identified at neighbourhood radius R=100 kpc, (b) for Sample1.2.

## 5. Summary

In this paper, we identify isolated Main galaxies by cluster analysis, and study the properties of isolated Main galaxy samples identified at different radii. In order to decrease the effect of radial selection function, we have constructed a subsample in the redshift region $0.08 \leq z \leq 0.12$, and refer to this subsample as "MGSU"(It contains 67777 galaxies). We set up the threshold that the number of isolated galaxy is below 10% of total galaxy number in the sample as the condition of producing isolated galaxy sample. Two isolated Main galaxy samples are analysed: Sample1.2, identified at radius r=1.2, includes 4099 galaxies (6.0% of the total galaxy number in "MGSU"); Sample1.4, identified at radius r=1.4, includes 2596 galaxies (3.8% of the total galaxy number in "MGSU"). It turns out that isolated Main galaxy samples identified at different radii have the same properties. This demonstrates that the cluster analysis and the threshold setting up by us are effective in the process of identifying isolated galaxies, the isolated galaxy sample identified below the threshold is a good one. Additionally, we also find that early-type galaxies in isolated Main galaxy samples are fewer than that in close double galaxy sample.